\documentclass[aps,pre,showpacs,amsmath,amssymb,amsfonts,superscriptaddress,twocolumn]{revtex4-1}

\usepackage{graphicx}
\usepackage{dcolumn}
\usepackage{bm}
\usepackage{epsf}
\usepackage{color}
\usepackage[colorlinks=true,citecolor=blue,linkcolor=blue,urlcolor=blue]{hyperref}

\newcommand{\mb}[1]{\mbox{\boldmath$#1$}}

\begin{document}

\title{Degenerate optimal paths in thermally isolated systems}

\author{Thiago V. Acconcia} 
\email[]{thiagova@ifi.unicamp.br}
\affiliation{Instituto de F\'isica `Gleb Wataghin', Universidade Estadual de Campinas, 13083-859, Campinas, S\~{a}o Paulo, Brazil}

\author{Marcus V. S. Bonan\c{c}a}
\email[]{mbonanca@ifi.unicamp.br}
\affiliation{Instituto de F\'isica `Gleb Wataghin', Universidade Estadual de Campinas, 13083-859, Campinas, S\~{a}o Paulo, Brazil}


\date{\today}

\begin{abstract} 
We present an analysis of the work performed on a system of interest that is kept thermally isolated during the switching of a control parameter. We show that there exists, for a certain class of systems, a finite-time family of switching protocols for which the work is equal to the quasistatic value. These optimal paths are obtained within linear response for systems initially prepared in a canonical distribution. According to our approach, such protocols are composed of a linear part plus a function which is odd with respect to time reversal. For systems with one degree of freedom, we claim that these optimal paths may also lead to the conservation of the corresponding adiabatic invariant. This points to an interesting connection between work and the conservation of the volume enclosed by the energy shell. To illustrate our findings, we solve analytically the harmonic oscillator and present numerical results for certain anharmonic examples.    
\end{abstract}

\maketitle
\section{Introduction \label{sec:intro}}

The accumulated and organized knowledge that we call \textit{thermodynamics} has been one of the main pillars of our physical understanding of the world around us. However, the only processes that are fully describable by means of classical thermodynamics are quasistatic ones, i.e., processes which are a succession of equilibrium states \cite{callen}. On the other hand, real thermodynamic processes happen in \textit{finite time} \citep{Sal1,Sal2,Sal3} and hence drive the system out of equilibrium. In this case, the second law imposes certain limits to the energy exchange between a system of interest and an external agent. How to get as close as possible to the minimal energetic cost of driving a system from one state to another in finite time remains then a crucial question. Thus, it is very desirable to develop a general method to solve such optimization problem. 

Thermodynamic processes can be performed under different constraints. The system of interest can be kept, for instance, in contact with  a heat bath during the time interval its externally controlled parameter is switched. In this situation, the minimal energetic cost is equal to the difference of Helmholtz free energies. Thereby, one of the many applications of such optimal finite-time processes is the estimation of free-energy differences \cite{watanabe,hunter,antonelli,zuckerman,lindberg,geiger}. A major breakthrough in this problem was achieved by Jarzynski \citep{Jarz0} and Crooks \cite{Crooks}. Through their results, finite-time processes can be used without leading to a biased estimation \cite{shirts,atilgan}. Nevertheless, one needs to sample extremely rare events in order to have reliable estimates \cite{Jarz1}. Besides, different demands for better efficiency in finite time have increased the interest on optimal control of thermodynamic systems \cite{esposito2,aurell,jordan,esposito,hoffman,xiao,zulkow,paolo,Sebastian}.

There are, at the moment, two main ways of finding optimal finite-time processes under isothermal conditions: stochastic models \cite{schmidel,gomez,engel} and linear response theory \cite{deKoning,Sivak,zulkowski,bonanca1}. The results obtained so far within the stochastic approach show intriguing, interesting, and not well understood features which appear only for sufficiently fast processes. On the other hand, the linear response approach provides an analytical treatment of a broader class of systems, although limited to quasiequilibrium processes. In the context of thermally isolated systems, the analysis of optimal paths has followed along the same lines as the stochastic approach for isothermal processes \cite{seifert}. However, an analytical description through stochastic methods is very restricted to linear systems and leaves open questions about what happens in the nonlinear case.

In the present work, we study the problem of finding optimal finite-time processes in thermally isolated systems via linear response theory. We focus on the regime in which the variation of the externally controlled parameter is small but has arbitrary speed. In Sec.~\ref{sec:excwork}, we derive an expression for the excess work \cite{Allah1,Allah2,Sivak}, i.e.,  a quantity that characterizes the energetic cost along a given process. In Secs.~\ref{sec:excoh} and \ref{sec:nonlisys}, this expression is employed to explain the numerical results of simple linear and nonlinear systems. The results obtained show unexpected features from the point of view of usual thermodynamic wisdom. In Sec.~\ref{sec:adinvar}, we connect the excess work to the adiabatic invariant, suggesting that every time the former vanishes the latter is conserved. We summarize and conclude in Secs.~\ref{sec:discus} and \ref{sec:conclu}.

\section{Thermally isolated systems and the excess work \label{sec:excwork}}

\begin{figure}
\centering
\includegraphics[scale = 0.4]{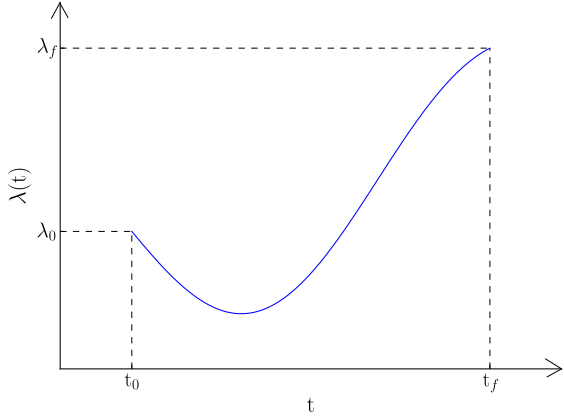} 
\caption{(Color online) Schematic representation of a given protocol $\lambda(t)$ performed by the external agent while the system is kept thermally isolated.}
\label{sketch_perturbation}
\end{figure}

Let us consider the following setup: first, we keep our system of interest in contact with a heat bath until its relaxation to the Boltzmann-Gibbs distribution,
\begin{equation}
\rho_{eq}(\mb{\Gamma}; \lambda_{0}) = \exp(-\beta \mathcal{H}(\mb{\Gamma}; \lambda_{0}))/ \mathcal{Z}(\beta, \lambda_{0})\,,
\label{exc0}
\end{equation} 
where $\mb{\Gamma}$ is a point in phase space, $\beta = (k_{B} T)^{-1}$, with $T$ and $k_{B}$ being the temperature of the heat bath and the Boltzmann constant, respectively. The quantity $\mathcal{Z}(\beta, \lambda_{0})$ denotes the partition function of the system given by $\mathcal{Z}(\beta, \lambda_{0}) = \int d\mb{\Gamma} \  \exp(-\beta \mathcal{H}(\mb{\Gamma}; \lambda_{0}))$ and $\lambda_{0}$ is the initial value of our control parameter $\lambda$. Second, the system is decoupled from the reservoir and kept thermally isolated while the external agent switches $\lambda$ from $\lambda_{0}$ to $\lambda_{f}$ according to a given protocol (see Fig. \ref{sketch_perturbation}). We express the protocol $\lambda(t)$ as follows:
\begin{equation}
\lambda(t) = \lambda_{0} + \delta\lambda \ g(t)\,,
\end{equation}  
where $g(t)$ is such that $g(t_{0}) = 0$ and $g(t_{f}) = 1$. Thus, the variation in $\lambda$ in the time interval $\tau \equiv t_{f}-t_{0}$ is 
$\delta\lambda = \lambda_{f} - \lambda_{0}$.

If we consider just a single realization of the protocol $\lambda(t)$, the probability of reaching the value $\mathcal{W}$ of the work performed is given by $\mathcal{P}(\mathcal{W}) = \langle \delta(\mathcal{W} - \mathcal{W}[\mb{\Gamma}_{t}])  \rangle$, where $\mb{\Gamma}_{t}$ is a given trajectory in phase space \cite{Cher}. This means that $\mathcal{P}(\mathcal{W})$ can be calculated from an average over all the possible trajectories or realizations \cite{Jarz0}. The thermodynamic work then reads
\begin{equation}
W = \int_{-\infty}^{\infty} d\mathcal{W}\,\mathcal{P}(\mathcal{W})\,\mathcal{W}\,,
\end{equation}
or, equivalently,
\begin{equation}
W = \int_{t_{0}}^{t_{f}} dt \, \frac{d \lambda}{dt}\, \overline{\dfrac{\partial \mathcal{H}}{\partial \lambda}} \,,
 \label{exc1}
\end{equation}
where $\bar{A}$ denotes the nonequilibrium average of the observable $A$.

Assuming that $|\delta\lambda\,g(t)/ \lambda_{0}| \ll 1$ for $t_{0} \leq t \leq t_{f}$, we can treat the effects of the generalized force $\partial \mathcal{H}/\partial\lambda$ perturbatively. In other words, once our system of interest is described by a Hamiltonian $\mathcal{H}[\lambda(t)]$, we can expand it in powers of $\delta\lambda$ as follows:
\begin{equation}
\mathcal{H}[\lambda(t)] = \mathcal{H}(\lambda_{0}) + \delta \lambda \ g(t) \dfrac{\partial \mathcal{H}}{\partial \lambda} + \mathcal{O}(2)\,.
\end{equation}
Therefore, the nonequilibrium average of the generalized force can be calculated by means of linear response theory \cite{kubo1,kubo2}. It reads
\begin{eqnarray}
\overline{\dfrac{\partial \mathcal{H}}{\partial \lambda}}(t) = \left\langle \dfrac{\partial \mathcal{H}}{\partial \lambda} \right\rangle_{0} 
+ \chi_{0}^{\infty} \delta\lambda g(t) - \delta\lambda \int_{t_{0}}^{t} ds \, \phi_{0}(t-s) \, g(s) ,\nonumber\\
\label{exc2}
\end{eqnarray}
where $\langle \cdot \rangle_{0}$ denotes an average on the initial ensemble, given by  Eq.~(\ref{exc0}), and the subscript refers to the value $\lambda_{0}$. The second term in the right-hand side of Eq.~(\ref{exc2}) describes the \textit{instantaneous} response, which is due to $\partial \mathcal{H}/\partial\lambda$ being a function of the external control $\lambda$ \cite{kubo1,kubo2}. In particular, we have 
\begin{equation}
\chi_{0}^{\infty} = \left\langle \frac{\partial^{2} \mathcal{H}}{\partial\lambda^{2}} \right\rangle_{0}\,.
\end{equation}
The second term describes the \textit{delayed} response and $\phi_{0}(t)$ is the so-called \textit{response} function. It will be convenient to express it in terms of the \textit{relaxation} function, $\Psi_{0}(t)$ \cite{kubo1,kubo2}. This can be done as follows:
\begin{equation}
\phi_{0}(t) = - \dfrac{d\Psi_{0}}{dt}(t) = - \beta \dfrac{d}{dt}(C_{0}(t) - \mathcal{C}) \,,
\label{respfunction}
\end{equation}
where $C_{0}(t) = \langle A(0) A(t) \rangle_{0}$ is the \textit{correlation} function of $A = \partial \mathcal{H}/\partial \lambda$ and the constant $\mathcal{C}$ is given by \cite{kubo2} 
\begin{equation}
\mathcal{C} = \lim_{\epsilon \to 0} \epsilon \int_{0}^{\infty} dt \ e^{-\epsilon t} \, C_{0}(t)\,. 
\label{constrelax}
\end{equation} 

Therefore, Eq.~(\ref{exc2}) can be rewritten after an integration by parts as 
\begin{eqnarray}
\overline{\dfrac{\partial \mathcal{H}}{\partial \lambda}}(t) &=& \left\langle \dfrac{\partial \mathcal{H}}{\partial \lambda} \right\rangle_{0} - \delta\lambda \tilde{\Psi}_{0}(0) g(t) \nonumber \\
&+& \delta \lambda \int_{0}^{t-t_{0}} du \, \Psi_{0}(u) \dfrac{dg}{dt'}\bigg|_{t'=t-u}\,,
\label{exc3}
\end{eqnarray}
where $\tilde{\Psi}_{0} \equiv \Psi_{0}(0) - \chi_{0}^{\infty}$. Finally, substituting Eq.~(\ref{exc3}) in expression (\ref{exc1}), we obtain 
\begin{eqnarray} \label{total_work}
W &=& \delta\lambda \left\langle\dfrac{\partial \mathcal{H}}{\partial\lambda} \right\rangle_{0} - \dfrac{(\delta \lambda)^{2}}{2} \tilde{\Psi}_{0}(0) \nonumber \\
&+&  (\delta \lambda)^{2}  \int_{t_{0}}^{t_{f}} dt \, \frac{dg}{dt} \int_{t_{0}}^{t} dt' \, \Psi_{0}(t-t') \frac{dg}{dt'}\,,
\end{eqnarray}
using the boundary conditions for $g(t)$. The first two terms of the previous expression \textit{do not} depend on the protocol $g(t)$. Indeed, it can be verified (see Appendix \ref{sec:apB}) that they are the first terms of the series expansion of the quasistatic work for $\delta\lambda/\lambda_{0} \ll 1$. The last term clearly depends on $g(t)$ and therefore represents the \textit{excess} work \cite{Allah1,Allah2,Sivak}. Since $\Psi_{0}(-t) = \Psi_{0}(t)$ [see Eq.~(\ref{respfunction})], we obtain 
\begin{eqnarray}\label{excess_work}
W_{exc} 
= \dfrac{(\delta \lambda)^{2}}{2}  \int_{0}^{1} du \int_{0}^{1} du' \, \dot{g}(u)\, \Psi_{0}(\tau(u-u')) \, \dot{g}(u')\,,\nonumber\\
\end{eqnarray}
where $\dot{g}(u)$ and $\dot{g}(u')$ denote the derivatives with respect to $u\equiv(t-t_{0})/\tau$ and $u'\equiv(t'-t_{0})/\tau$, respectively.

In summary, linear response expresses the total work as a sum of two contributions. One is independent of the particular process and is
identical to what would be obtained in the quasistatic limit, i.e., the quasistatic work. The other is therefore interpreted as the additional amount of energy that the external agent has to pump into the system in a finite-time process. What we call excess work here is therefore defined as $W_{exc} \equiv W - W_{qs}$, where $W_{qs}$ is the quasistatic work. Thus, we expect that Eq.~(\ref{excess_work}) goes to zero asymptotically as the quasistatic limit is approached.

\section{Excess work for an exactly solvable model \label{sec:excoh}}

We shall apply now the expression of Eq. (\ref{excess_work}) to the one-dimensional harmonic oscillator, which is a completely solvable model that allows us to check the accuracy of the linear response expression of the excess work. Therefore, we will consider that the dynamics of our system of interest is given by the following time-dependent Hamiltonian:
\begin{equation}
\mathcal{H}[\lambda(t)] = \dfrac{p^{2}}{2} + \lambda(t) \dfrac{x^{2}}{2}\,,
\label{ohhamil}
\end{equation} 
which can model, for example, the motion of a colloidal particle in an optical trap \cite{evans}. 

From the solution of Hamilton's equations for $\lambda = \lambda_{0}$, we can calculate the relaxation function exactly. According to Eq.~(\ref{respfunction}), we first need to obtain the correlation function $C_{0}(t)$,
\begin{equation}
C_{0}(t) = \left\langle \dfrac{x^{2}(0)}{2} \dfrac{x^{2}(t)}{2} \right\rangle_{0} = \dfrac{2 \cos^{2}(\omega_{0}t) + 1}{4 \beta^{2}\lambda_{0}^{2}} \,,
\label{correfoh}
\end{equation}
where $\omega_{0}^{2} \equiv \lambda_{0}$ since the mass was set equal to one in Eq. (\ref{ohhamil}). The next step is to calculate the constant $\mathcal{C}$. From Eqs.~(\ref{constrelax}) and (\ref{correfoh}), we obtain $\mathcal{C} = 2/(4 \beta^{2} \lambda_{0}^{2})$. Finally, the relaxation function reads 
\begin{equation}
\Psi_{0}(t) = \dfrac{\cos(2\omega_{0}t)}{4\beta\lambda_{0}^{2}} = \Psi_{0}(0) \cos(2\omega_{0}t)\,,
\label{relaxfoh}
\end{equation}
where $\Psi_{0}(0) \equiv (\beta/2) (\langle x^{4}(0)/4 \rangle_{0} - \langle x^{2}(0)/2 \rangle_{0}^{2})$. Although it is a bit misleading to call Eq.~(\ref{relaxfoh}) a relaxation function, we will see next that it leads to a reasonable thermodynamic behavior of $W_{exc}$ \citep{Allah1,Allah2}.
	
\begin{figure}
\includegraphics[scale = 0.60]{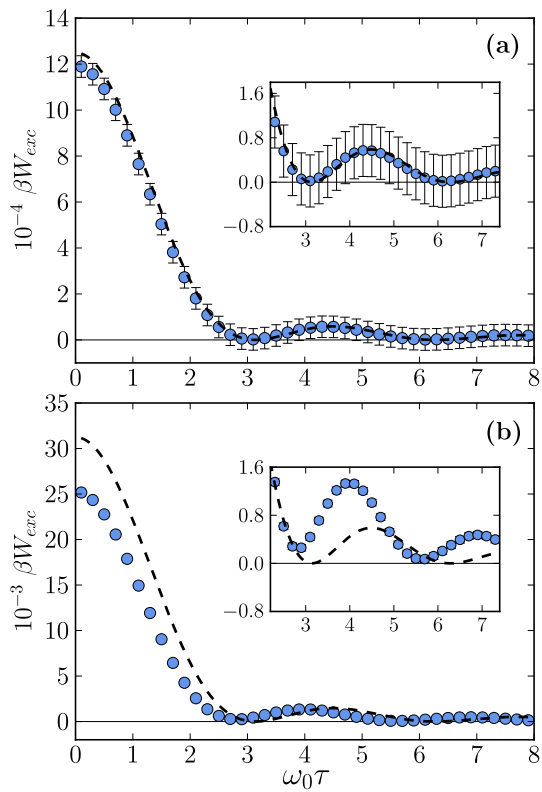} 
\caption{(Color online) Comparison between numerical calculations (blue circles) and Eq.~(\ref{excworkoh}) (dashed line) for (a) $\delta\lambda/\lambda_{0} = 0.1$ and (b) $\delta\lambda/\lambda_{0} = 0.5$. We used $10^{6}$ initial conditions to calculate the work for each value of the switching time $\tau$.}
\label{fig_harm_osc}
\end{figure}

Substituting Eq.~(\ref{relaxfoh}) into (\ref{excess_work}) and using the linear protocol $g(t) = (t-t_{0})/\tau$, we obtain
\begin{equation}
W_{exc} (\tau) = \dfrac{1}{8\beta} \left(\dfrac{\delta\lambda}{\lambda_{0}}\right)^{2} \dfrac{\sin^{2}(\omega_{0} \tau)}{(\omega_{0}\tau)^{2}}\,.
\label{excworkoh}
\end{equation}
This expression goes to zero in the quasistatic limit, $\tau\to\infty$, and has its maximum value when $\tau\to 0$. Figure \ref{fig_harm_osc} shows a comparison between Eq.~(\ref{excworkoh}) and numerical simulations. The agreement is very good for $\delta\lambda/\lambda_{0} = 0.1$. Nevertheless, our linear response expression already deviates considerably for $\delta\lambda/\lambda_{0} = 0.5$.

A very striking prediction of Eq.~(\ref{excworkoh}) is that $W_{exc}$ can be zero for specific \textit{finite} values of $\tau$. In other words, there are finite switching times for which the total work is equal to the quasistatic value. These particular values of $\tau$ can be obtained directly from Eq. (\ref{excworkoh}): whenever $\omega_{0}\tau = l\pi$, with $l$ integer, we have $W_{exc}=0$. This means that already for $\tau$ equal to half of the natural period of oscillations, $2\pi/\omega_{0}$, the system can be driven as if the process was a quasistatic one. 
The possibility of achieving the quasistatic value of the work performed in a finite-time process was pointed out before in Ref.~\cite{seifert}, though without addressing the dependence of the excess work on the switching time. In Fig.~\ref{fig_harm_osc}, the numerical value of $W_{exc}$ is obtained after subtracting from $W$ the exact value of the quasistatic work $W_{qs}$,
\begin{equation}
W_{qs} = \dfrac{1}{\beta}\left[ \left(\dfrac{\lambda_{f}}{\lambda_{0}} \right)^{1/2} - 1 \right]\,,
\end{equation}
More details about the previous expression for $W_{qs}$ can be found in Appendix \ref{sec:apB}. Finally, we point out that the inset of Fig.~\ref{fig_harm_osc}(b) indicates that the minima of $W_{exc}$ are lifted as $\delta\lambda/\lambda_{0}$ increases. This effect is clearly beyond our linear response approach.

\begin{figure}
\includegraphics[scale = 0.60]{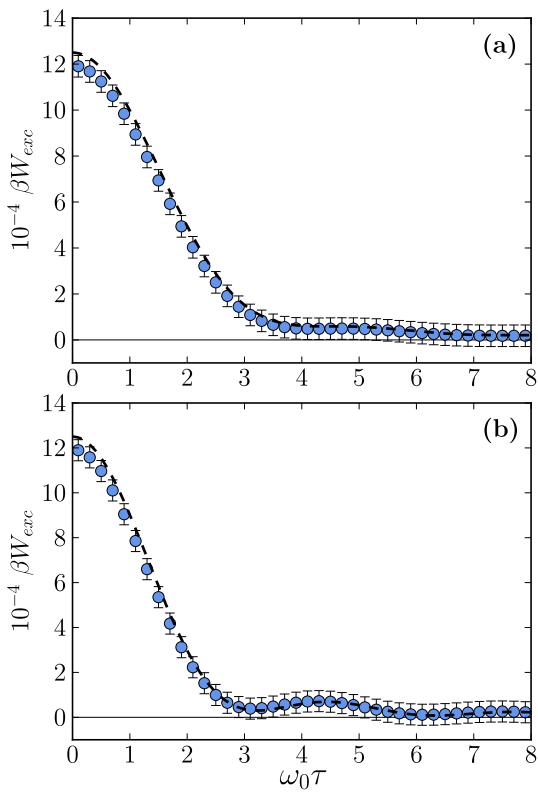} 
\caption{(Color online) Comparison between numerical calculations (dashed line) and Eq.~(\ref{excworkoh}) (blue circles) for the (a) quadratic, $g(t) = (t-t_{0}/\tau)^{2}$, and the (b) exponential, $g(t) = (1 - e^{-(t-t_{0})/\tau}) / (1-e^{-1})$, protocols. In both cases, $\delta\lambda/\lambda_{0} = 0.1$.  We used $10^{6}$ initial conditions to calculate the work for each value of the switching time $\tau$.}
\label{quadratic_expo}
\end{figure}

These results tell us that a simple linear protocol can be optimal if we choose the value of $\tau$ carefully. Besides, our linear response expression predicts that this happens only for specific finite values of $\tau$. Among the many interesting questions that arise from these remarks, we will focus now on the following: Is this a special feature of the linear driving? To investigate that, we will compare analytical and numerical results for different nonlinear protocols. Let us consider, for instance, a quadratic, $g(t) = ((t-t_{0})/\tau)^{2}$, and an exponential, $g(t) = (1 - e^{-(t-t_{0})/\tau}) / (1-e^{-1})$, protocol. The results are shown in Fig.~\ref{quadratic_expo}. The agreement is again very good and $W_{exc}$ goes to zero in the quasistatic limit. In contrast to what happens for the linear driving, our analytical results predict that $W_{exc}$ \textit{never} vanishes in finite time for the nonlinear protocols considered. This can be explicitly checked, for instance, for the quadratic protocol, whose expression for $W_{exc}$ reads
\begin{eqnarray}
W_{exc} &=&\frac{1}{8\beta}\left(\frac{\delta\lambda}{\lambda_{0}}\right)^{2}\nonumber\\
&\times & \frac{\left[ (\omega_{0}\tau)^{2}-\omega_{0}\tau\,\sin{(2\omega_{0}\tau)}+\sin^{2}{(\omega_{0}\tau)}\right]}{(\omega_{0}\tau)^{4}}\,.
\end{eqnarray}
Nevertheless, Fig. \ref{quadratic_expo} shows that $W_{exc}$ does have finite-time minima in these cases.

It has been shown in the literature \cite{seifert} that there exists indeed a highly degenerate family of finite-time nonlinear protocols for which the work performed is equal to the quasistatic one. In what follows, we will show how to obtain such protocols analytically from our linear response approach.

\begin{figure}
\includegraphics[scale = 0.40]{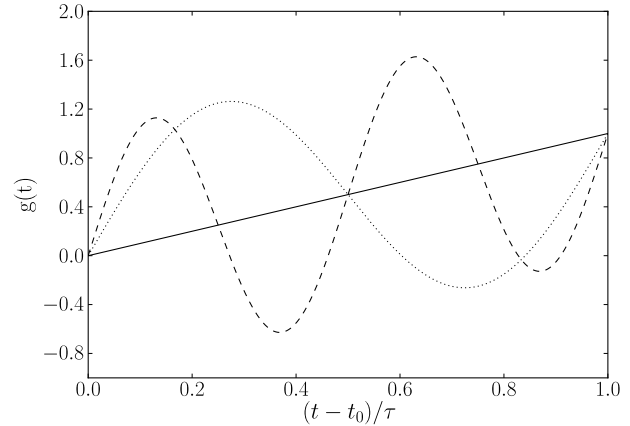} 
\caption{Examples of the family of protocols given by Eq.~(\ref{gen_protocol}) for $\kappa=0$ (solid line), $\kappa = 4$ (dashed line), and $\kappa=2$ (dotted line). We fixed $a$ equal to one.}
\label{fig_protocols}
\end{figure} 

First, we use the fact that, for $g(t)=(t-t_{0})/\tau$, we do observe zeros of $W_{exc}$ in finite time. Second, it can be easily verified that the Fourier series of the linear protocol has no cosine coefficients in the interval  $t_{0} \leq t \leq t_{f}$. Therefore, we wonder what happens to $W_{exc}$ if we perform a protocol given by a linear part plus a sine function such that the boundary conditions $g(t_{0})=0$ and $g(t_{f})=1$ are preserved. In other words, we ask ourselves whether the protocol  
\begin{eqnarray}\label{gen_protocol}
g(t) = \dfrac{t-t_{0}}{\tau} + a \sin\left(\kappa \pi \dfrac{(t-t_{0})}{\tau}\right) \,,
\end{eqnarray}
where $\kappa$ is an integer and $a$ is an arbitrary real number, leads to zeros of $W_{exc}$. Some examples of these functions can be seen in Fig.~\ref{fig_protocols}.

\begin{figure}
\includegraphics[scale = 0.60]{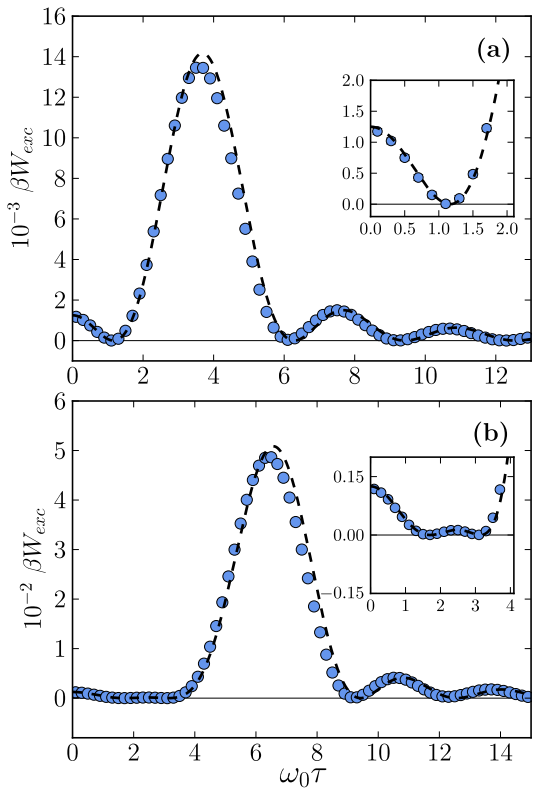} 
\caption{(Color online) $W_{exc}$ for the protocols (a) $g(t) = (t-t_{0})/\tau + \sin(2\pi (t-t_{0})/\tau)$ and (b) $g(t) = (t-t_{0})/\tau + \sin(4\pi (t-t_{0})/\tau)$. Analytical and numerical results are represented by dashed lines and blue circles, respectively. We used $\delta\lambda/\lambda_{0} = 0.1$ and $10^{6}$ initial conditions. }
\label{sin_2x}
\end{figure} 	

We show in Fig.~\ref{sin_2x} the comparison between analytical and numerical calculation of $W_{exc}$ using the protocol given by Eq. (\ref{gen_protocol}). These results show two additional features compared to those in Fig.~\ref{fig_harm_osc}: the first zero of $W_{exc}$ occurs at shorter times and there is a constructive resonance for a specific value of $\tau$. These features can be better explained if, using Eq.~(\ref{relaxfoh}), we rewrite Eq.~(\ref{excess_work}) as
\begin{eqnarray}
\lefteqn{W_{exc}\propto}\nonumber\\
&  & \left(\int_{0}^{1} du \, \dot{g}(u) \cos{(2\omega_{0}\tau u)}\right)^{2} 
+ \left(\int_{0}^{1} du \, \dot{g}(u) \sin(2\omega_{0}\tau u)\right)^{2} .\nonumber\\
\end{eqnarray}
Therefore, to have $W_{exc}=0$, we demand that
\begin{subequations}
\begin{align}
\int_{0}^{1} du \, \dot{g}(u)\, \cos{(2\omega_{0}\tau u)} &= 0 \,,\\
\int_{0}^{1} du \, \dot{g}(u)\, \sin{(2\omega_{0}\tau u)} &= 0 \,,
\end{align}
or, equivalently,
\begin{align}
z \equiv \int_{0}^{1} du \, \dot{g}(u) \ e^{2i\omega_{0}\tau u} = 0\,.
\label{eq_z}
\end{align}
\end{subequations}
By inserting the protocol of Eq. (\ref{gen_protocol}) in the expression for $z$ and solving the integral, we obtain
\begin{eqnarray}
z &=& 2 i \omega_{0} \tau \left[ \dfrac{1- \cos(2\omega_{0}\tau)}{(2\omega_{0}\tau)^{2}} + a\pi \kappa \dfrac{[1-\cos(\kappa\pi)\cos(2\omega_{0}\tau)]}{(2\omega_{0}\tau)^{2} - (\kappa\pi)^{2}}\right]  \nonumber \\
&+& 2 \omega_{0} \tau \left[\dfrac{\sin(2\omega_{0}\tau)}{(2\omega_{0}\tau)^{2}} + a\pi \kappa \dfrac{\cos(\kappa\pi)\sin(2\omega_{0}\tau)}{(2\omega_{0}\tau)^{2} - (\kappa\pi)^{2}}  \right]\,.
\end{eqnarray}

\begin{figure}
\includegraphics[scale = 0.60]{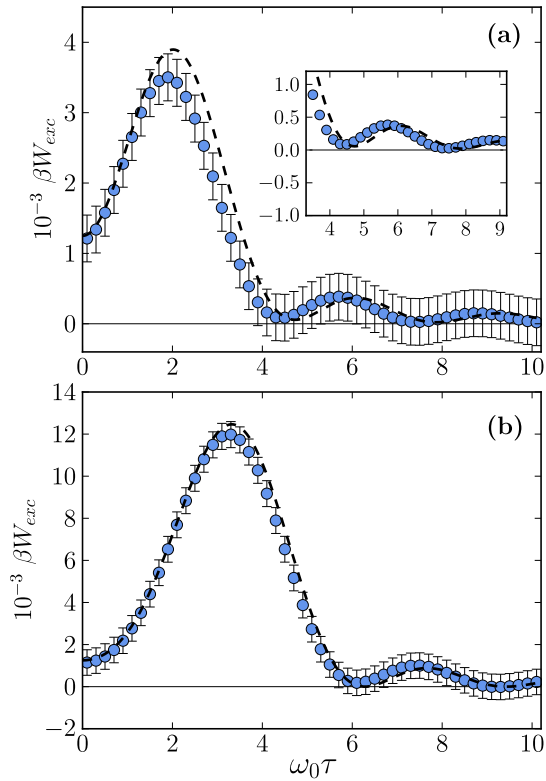} 
\caption{(Color online) $W_{exc}$ for the protocol (\ref{gen_protocol}) with (a) $\kappa = 1$ and $a = 1$ and (b) $\kappa = 2$ and $a = -1$. Analytical and numerical results are represented by dashed lines and blue circles, respectively. We used $\delta\lambda/\lambda_{0} = 0.1$ and $10^{5}$ initial conditions. }
\label{a_negative}
\end{figure}

We conclude that the real and imaginary parts of $z$ can be zero simultaneously if $\kappa$ is \textit{even} and $2\omega_{0}\tau = 2\pi l$, with $l$  integer, \textit{independently} of the value of $a$. This condition predicts that the first zero would occur for $\omega_{0}\tau = \pi$ unless $\kappa=2$. However, we see in Fig.~\ref{sin_2x} that the position of the very first minima of $W_{exc}$ does not follow this prediction. This is so because there is a second kind of zero that does depend on the value of $a$. As before, we demand that real and imaginary parts of $z$ vanish, but now for the \textit{same} value of $a$. We obtain then
\begin{equation}
\omega_{0}\tau = \frac{(\kappa\pi/2)}{(1 + \kappa\pi a)^{1/2}}\,.
\label{eq_a}
\end{equation} 
For $\kappa=2$ and $a=1$, Eq.~(\ref{eq_a}) leads to $\omega_{0}\tau\approx 1.2$, in agreement with Fig.~\ref{sin_2x}(a). For $\kappa=4$ and $a=1$, there are two zeros before the resonant peak, with the first one at $\omega_{0}\tau \approx 1.7$, due to the value of $a$, and the second one at $\omega_{0}\tau = \pi$. Hence, we can obtain zeros of $W_{exc}$ at arbitrarily short times by choosing the values of $a$ appropriately. Nevertheless, for $a$ negative, we are limited by the square root in Eq.~(\ref{eq_a}). Figure \ref{a_negative}(b) shows what happens to $W_{exc}$ when we perform the protocol of Eq. (\ref{gen_protocol}) with $a =-1$ and $\kappa = 2$. As opposed to Fig.~\ref{sin_2x}(a), the zeros of $W_{exc}$ in Fig.~\ref{a_negative}(b) do not depend on $a$. Although the results in Fig.~\ref{a_negative}(a) show very pronounced minima, linear response predicts that there are no finite-time zeros of $W_{exc}$ in this case. 

\begin{figure}
\includegraphics[scale = 0.40]{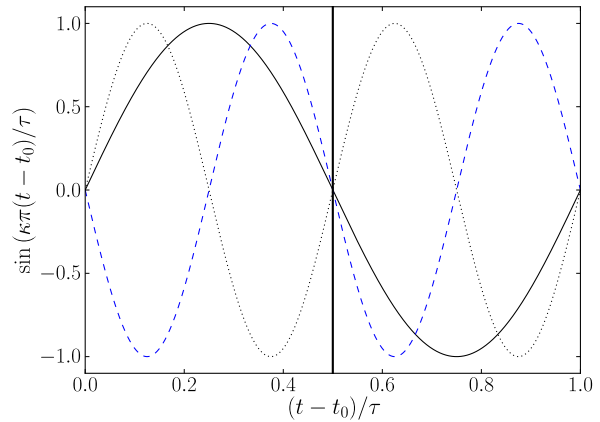} 
\caption{(Color online) Sinusoidal terms, $\sin{\left(\kappa\pi(t-t_{0})/\tau\right)}$, for $\kappa=2$ (solid line) and $\kappa=4$ (dotted line), and $-\sin{\left(\kappa\pi(t-t_{0})/\tau\right)}$ for $\kappa=4$ (dashed line).}
\label{symmetry_sine_protocols}
\end{figure}

\begin{figure}
\includegraphics[scale = 0.40]{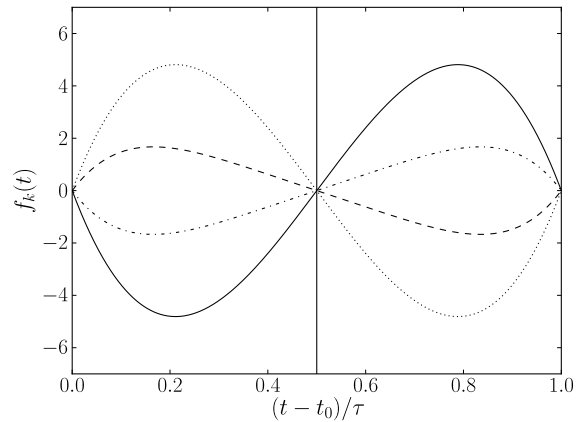} 
\caption{Examples of the polynomials given by Eq.~(\ref{polynomials}) for $k=1$ and $\epsilon=+1$ (dotted line), $k=2$ and $\epsilon=+1$ (dashed line), $k=1$ and $\epsilon=-1$ (solid line), and $k=2$ and $\epsilon=-1$ (dash-dotted line).}
\label{symmetry_polynomials_protocols}
\end{figure}

This analysis of $W_{exc}$ has interesting consequences. If we add to the protocol presented in Eq. (\ref{gen_protocol}) an arbitrary number of sine functions with arbitrary coefficients and even values of $\kappa$, we still obtain zeros when $\omega_{0}\tau = l \pi$. This sum of sinusoidal terms can be understood as the Fourier series of a function whose values at $t=t_{0}$ and $t=t_{f}$ are zero and which is \textit{odd} with respect to a change of $(t-t_{0})$ by $\tau-(t-t_{0})$. This property is illustrated in Fig.~\ref{symmetry_sine_protocols}. Therefore, we conclude that \textit{any} function that vanishes at $t_{0}$ and $t_{f}$ and is odd with respect to time reversal leads to the above-mentioned zeros of $W_{exc}$ when added to the linear protocol. For instance, the following family of polynomials:
\begin{equation}\label{polynomials}
f_{k}(t) = \epsilon\left[\dfrac{1-2((t-t_{0})/\tau)}{2^{2k+1}} + \left(\dfrac{t-t_{0}}{\tau}-\dfrac{1}{2}\right)^{2k+1} \right],
\end{equation}
where $k$ is an integer and $\epsilon = \pm 1$, has such property, as illustrated in Fig.~\ref{symmetry_polynomials_protocols}. For $\kappa = 1$, the sinusoidal term of Eq.~(\ref{gen_protocol}) is not odd with respect to time reversal. 

%

\section{Nonlinear systems \label{sec:nonlisys}}

\begin{figure}
\includegraphics[scale=0.4]{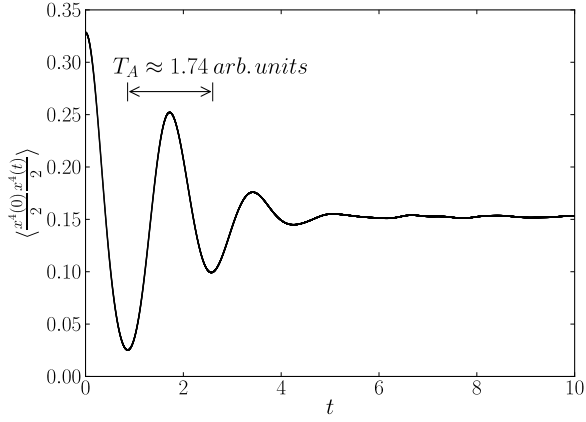}
\caption{Numerical calculation of $\langle x^{4}(0) x^{4}(t) \rangle_{0}/4$ for the oscillator (\ref{quartic}) with time-independent $\lambda$ equal to $\lambda_{0}$. Initial conditions were sampled according to a Boltzmann-Gibbs distribution with $\lambda_{0} = 1$. The period of oscillations $T_{A}$ is approximately equal to $1.74$ in arbitrary units. We used $10^{6}$ initial conditions.}
\label{correfunq}
\end{figure}

The linear response expression (\ref{excess_work}) depends strongly on the behavior of a specific autocorrelation function. For the system (\ref{ohhamil}), this function oscillates indefinitely when the dynamics is time independent. This is a special feature of linear systems whose observables have frequencies of oscillations which are independent of the energy. In contrast, the dynamics of nonlinear systems, as 
\begin{subequations}
\begin{align}
\label{quartic}
\mathcal{H}[\lambda(t)] = \dfrac{p^{2}}{2} + \lambda(t) \dfrac{x^{4}}{2}\,,
\end{align}
and 
\begin{align}\label{sixth}
\mathcal{H}[\lambda(t)] = \dfrac{p^{2}}{2} + \lambda(t) \dfrac{x^{6}}{6}\,,
\end{align}
\end{subequations}
present frequencies of oscillations which are energy dependent. Thus, correlation functions are oscillatory only when initial conditions are sampled from a single energy shell. Otherwise, a decay is observed due to the incommensurability of the superposed oscillations from different energy shells. Figure (\ref{correfunq}) shows an example of this for the system given in Eq. (\ref{quartic}).

In this section, we argue that the previous analysis of $W_{exc}$ can be extended to one-dimensional anharmonic oscillators. In particular, we want to investigate the behavior of $W_{exc}$ when systems (\ref{quartic}) and (\ref{sixth}) are driven by the protocols discussed in the previous section. As shown in Fig.~ \ref{correfunq}, the relevant correlation function of system (\ref{quartic}) has well-defined oscillations for short times. For the sake of clarity, let us assume for a moment that these oscillations last indefinitely with a period $T_{A}$. In this case, we can replace the actual $\Psi_{0}(t)$ by its Fourier series in the interval $[-T_{A}/2,T_{A}/2]$,
\begin{equation}\label{gen_psi}
\Psi_{0}(t)  =  \Psi_{0}(0) \sum_{n=1}^{\infty} a_{n} \cos(n \omega_{A} t)\,,
\end{equation}	
recalling that $\Psi_{0}(-t) = \Psi_{0}(t)$. The $a_{n}$ are the Fourier coefficients and $\omega_{A}\equiv 2\pi/T_{A}$ [according to Eq.~(\ref{respfunction}), $\Psi_{0}(t)$ would oscillate around zero implying that the coefficient $a_{0}$ is zero]. Substituting the previous expression into Eq.~(\ref{excess_work}), we have
\begin{eqnarray}\label{excess_gen_psi}
W_{exc} &=&  \dfrac{(\delta\lambda)^{2}}{2}\Psi_{0}(0)\nonumber\\  
&\times &\sum_{n=1}^{\infty} a_{k} \int_{0}^{1} du   \int_{0}^{1} du'\, \dot{g}(u) \cos(n \omega_{A}\tau(u-u')) \dot{g}(u').\nonumber\\
\end{eqnarray}

Analogously to Sec. \ref{sec:excoh}, Eq.~(\ref{excess_gen_psi}) can be written as
\begin{eqnarray}\label{excess_gen_psi2}
W_{exc} &=&  \sum_{n=1}^{\infty} A_{n} \left[ \left( \int_{0}^{1} du \ \dot{g}(u) \cos(n \omega_{A} \tau u) \right)^{2} \right. \nonumber \\
&+& \left. \left( \int_{0}^{1} du \ \dot{g}(u) \sin(n \omega_{A} \tau u) \right)^{2} \,\right] \,,
\end{eqnarray}
where $A_{n} = ((\delta\lambda)^{2}/2) \Psi_{0}(0) \ a_{n}$. This expression for the excess work vanishes if, for instance, each term of the sum is zero for the same value of $\tau$. To verify this possibility, we check under what conditions
\begin{equation}
z_{A} \equiv \int_{0}^{1} du \ \dot{g}(u) \ e^{i n\omega_{A}\tau u} = 0\,.
\end{equation}	

Using the protocol of Eq. (\ref{gen_protocol}), the quantity $z_{A}$ reads
\begin{eqnarray}\label{expza}
\lefteqn{z_{A} =}\nonumber\\ 
&i n \omega_{A}& \tau \left[ \dfrac{1- \cos(n\omega_{A}\tau)}{(n\omega_{A}\tau)^{2}} + a\pi \kappa \dfrac{(1-\cos(\kappa\pi)\cos(n\omega_{A}\tau))}{(n\omega_{A}\tau)^{2} - (\kappa\pi)^{2}}\right]  \nonumber \\
&+& n\omega_{A} \tau \left[\dfrac{\sin(n\omega_{A}\tau)}{(n\omega_{A}\tau)^{2}} + a\pi \kappa \dfrac{\cos(\kappa\pi)\sin(n\omega_{A}\tau)}{(n\omega_{A}\tau)^{2} - (\kappa\pi)^{2}}  \right]\,.
\end{eqnarray}
Thus, for $\kappa$ even, there are zeros of $z_{A}$ whenever $n\omega_{A}\tau = 2\pi l_{n}$, with $l_{n}$ an integer, except for $2l_{n}=\kappa$. Therefore, for $\kappa = 0$, the smallest value of $\omega_{A}\tau$ that provides a zero of $z_{A}$ for all modes simultaneously is $2\pi$. When $\kappa = 2$, this zero is forbidden by the denominators in Eq.~(\ref{expza}) and $\omega_{A} \tau = \pi$ is the first zero.

\begin{figure}
\includegraphics[scale = 0.60]{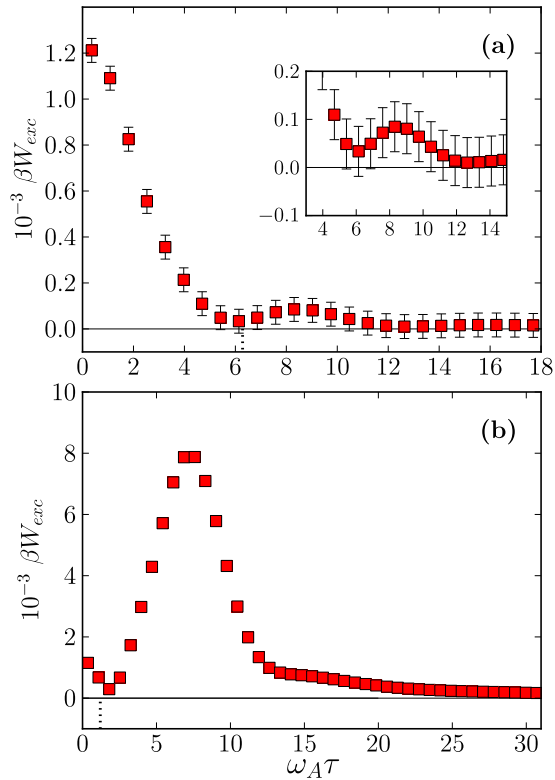} 
\caption{(Color online) Numerical calculation of $W_{exc}$ for the anharmonic oscillator (\ref{quartic}) using (a) a linear protocol and (b) the protocol (\ref{gen_protocol}) with $\kappa = 2$ and $a = 1$. In both cases, $\delta\lambda/\lambda_{0} = 0.1$ and we used $10^{6}$ initial conditions. The vertical dotted lines indicate the analytical prediction, (a) $\omega_{A}\tau = 2\pi$ and (b) $\omega_{A}\tau\approx 1.2$, of the first minimum.}
\label{fig:wexc_quartic}
\end{figure}

This prediction of the first minimum of $W_{exc}$ is in very good agreement with the numerical calculations shown in Fig.~\ref{fig:wexc_quartic}. However, it is based on a wrong assumption about the behavior of $\Psi_{0}(t)$. Introducing a small damping of oscillations, the Fourier transform of $\Psi_{0}(t)$ changes from a delta-like peak at $\omega_{A}$ to a peak with a small width whose position is very close (but not exactly equal) to $\omega_{A}$. Therefore, instead of the representation given by Eq. (\ref{gen_psi}), we would have the following one:
\begin{equation}\label{psifourier}
\Psi_{0}(t) = \Psi_{0}(0) \sqrt{\frac{2}{\pi}} \int_{0}^{\infty} d\omega\,\chi(\omega) \cos{(\omega t})\,,
\end{equation}
where $\chi(\omega)$ is the cosine Fourier transform of $\Psi_{0}(t)/\Psi_{0}(0)$. It is not hard to see that by plugging the expression (\ref{psifourier}) into Eq.~(\ref{excess_work}), we obtain an expression similar to (\ref{excess_gen_psi2}) with the sum replaced by the integral over $\omega$ and the coefficient $A_{n}$ replaced by $(\sqrt{2/\pi})\chi(\omega)$. In this case, we have a continuous (but small) interval of frequencies that contribute to $W_{exc}$ instead of the discrete values $n\omega_{A}$. Then, we can think of a $z_{A}(\omega)$ given exactly by Eq.~(\ref{expza}) with $n\omega_{A}$ replaced by $\omega$. To observe a zero in the excess work, $z_{A}(\omega)$ would have to be zero for all $\omega$ in a small vicinity of $\tilde{\omega}_{A}=\omega_{A}+\delta$ ($\delta$ represents the small shift of the peak due to damping). However, due to the incommensurability of such  frequencies, if $z_{A}(\tilde{\omega}_{A})$ is exactly zero, it is certainly nonzero for $\omega$ around $\tilde{\omega}_{A}$. In other words, if $z_{A}(\tilde{\omega}_{A}) = 0$, then $z_{A}(\omega) \approx 0$ for $\omega\approx\tilde{\omega}_{A}$, and a minimum of $W_{exc}$ arises. Therefore, our linear response approach predicts that $W_{exc}$ does not vanish in finite time for the family of protocols (\ref{gen_protocol}) when $\Psi_{0}(t)$ has damped oscillations. 

\begin{figure}
\includegraphics[scale = 0.40]{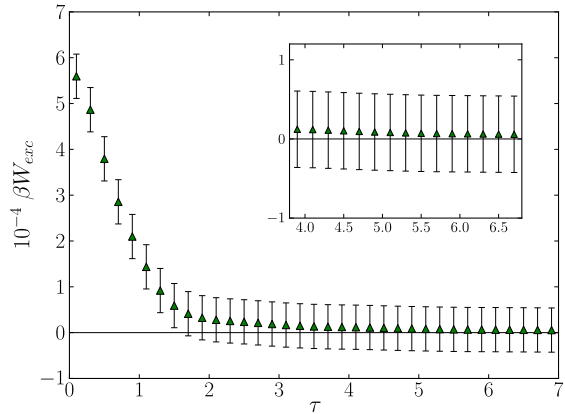} 
\caption{(Color online) Numerical calculation of $W_{exc}$ for the anharmonic oscillator (\ref{sixth}) using the linear protocol and $\delta\lambda/\lambda_{0} = 0.1$. We used $10^{6}$ initial conditions. The inset shows a small region around $\tau = 5.2$.}
\label{wexcess_sixth}
\end{figure}

As mentioned before, Fig. \ref{fig:wexc_quartic} illustrates the preceding discussion. The value of $W_{qs}$ was obtained analytically in Appendix~\ref{sec:apB}. Figure \ref{fig:wexc_quartic}(a) shows a numerical calculation of $W_{exc}$ for the system given by expression (\ref{quartic}) using the linear protocol. We kept $\delta\lambda/\lambda_{0} = 0.1$ so that our approach based on linear response theory is still valid. This numerical result shows that $W_{exc}$ has indeed minima and the position of the first one is very close to $\omega_{A}\tau =2\pi$. The numerical calculation of $W_{exc}$ for the protocol given by Eq. (\ref{gen_protocol}) is shown in Fig.~\ref{fig:wexc_quartic}(b). It shows the same features we observe in Fig.~\ref{sin_2x} for the harmonic oscillator, including a minimum at very short time scales for $\omega_{A}\tau\approx 1.2$. As discussed in Sec.~\ref{sec:excoh}, this minimum is related to the value of $a$. For the system (\ref{sixth}), the numerical result is shown in Fig.~\ref{wexcess_sixth} using the linear protocol. Although the excess work also approaches zero as for a finite $\tau$, we cannot distinguish between a minimum and a monotonic decay.

\begin{figure}
\includegraphics[scale = 0.40]{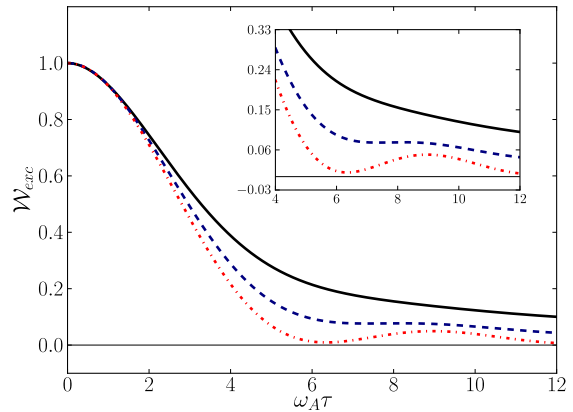} 
\caption{(Color online) Excess work, $\mathcal{W}_{exc} = 2 W_{exc}/ ((\delta\lambda)^{2} \Psi_{0}(0))$, obtained from expression (\ref{exponential_relaxation_func}) for $\alpha/\omega_{A} = 0.01$ (red dotted line), $0.1$ (blue dashed line), and $0.3$ (black solid line) using the linear protocol.}
\label{exp_relaxation}
\end{figure}

This analysis of $W_{exc}$ for the system (\ref{quartic}) shows how determinant is the behavior of the relaxation function. If correlations decay sufficiently fast, a crossover is observed in the behavior of $W_{exc}$. This can be easily verified using the phenomenological expression
\begin{equation}\label{exponential_relaxation_func}
\Psi(t) = \Psi_{0}(0) \,e^{-\alpha \vert t \vert} \left(\cos(\omega_{A} t) + \dfrac{\alpha}{\omega_{A}} \sin(\omega_{A} \vert t \vert) \right)\,,
\end{equation}
where $\alpha$ and $\omega_{A}$ denote the decay rate and the frequency of oscillations, respectively. The excess work obtained for the linear protocol using Eqs.~(\ref{exponential_relaxation_func}) and (\ref{excess_work}) is shown in Fig.~\ref{exp_relaxation} for different ratios of $\alpha/\omega_{A}$. As the decay rate increases, the minima disappear and $W_{exc}$ varies monotonically with $\tau$. We have verified the same sort of crossover when correlations decay as a power law.

%
	
\section{Excess work and the adiabatic invariant \label{sec:adinvar}}

Motivated by the results of the previous sections, we argue here how the excess work is connected to an important quantity of time-dependent Hamiltonian systems, namely, the adiabatic invariant. Adiabatic invariants are approximate constants of motion of time-dependent systems perturbed by slowly varying parameters \cite{hertz,Fasano}. Generally, the perfect conservation of an adiabatic invariant $\Omega$ is reached only in the quasistatic limit. Besides, for systems with one degree of freedom, it is possible to estimate analytically the conservation of $\Omega$ with high accuracy \cite{robnik2}. For this class of systems, the results obtained previously show that it is indeed possible to have $W$ approximately or even exactly equal to $W_{qs}$ for finite values of switching time. Thus, it is reasonable to expect that the $\Omega$ is conserved (or almost conserved) at exactly those values of $\tau$ for which $W_{exc}$ vanishes (or has minima). 

\begin{figure}
\includegraphics[scale = 0.40]{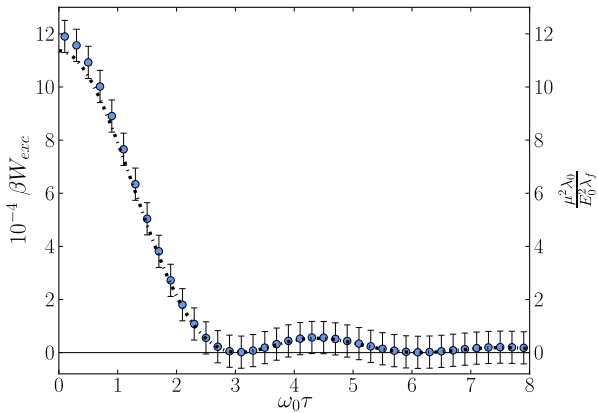} 
\caption{(Color online) Numerical calculations of $W_{exc}$ (blue dots) and $\mu^{2}$ (dashed line) for the harmonic oscillator (\ref{ohhamil}) using the linear protocol. We fixed $\delta\lambda/\lambda_{0} = 0.1$ and used $10^{6}$ initial conditions.}
\label{inv_adiabatic_harm} 
\end{figure}	

In what follows, we investigate the behavior of $\Delta \Omega\equiv \Omega(t_{f})-\Omega(t_{0})$ for systems given by expressions (\ref{ohhamil}) and (\ref{quartic}). The adiabatic invariant $\Omega$ of both oscillators is equal to the area enclosed by the energy shell and, according to Appendix \ref{sec:apC}, it can be expressed in terms of the energy and $\lambda$ only. Therefore, for fixed values of $\lambda_{0}$ and $\lambda_{f}$, $\Delta \Omega$ depends exclusively on the initial and final energies as we change $\tau$. It is shown in Ref.~\cite{robnik2} that if we sample initial conditions with the same value of energy, say $E_{0}$, and evolve the equations of motion of system (\ref{ohhamil}) for an arbitrary time interval $\tau$ using a given protocol $\lambda(t)$, the distribution of final energies $E_{1}$ is such that 
\begin{subequations}\label{energy}
\begin{align}
\overline{(E_{1} - \overline{E_{1}})^{2m+1}}\; =&\; 0\,, \\ 
\overline{(E_{1} - \overline{E_{1}})^{2m}}\; =&\; \dfrac{(2m+1)!!}{m!} \left(\overline{(E_{1} - \overline{E_{1}})^{2}}\right)^{m}\,,
\end{align}
\end{subequations}
where $m$ is an integer and the overbar denotes an average on the distribution of $E_{1}$. Equations (\ref{energy}) show that any moment of this distribution can be written in terms of the averaged energy $\overline{E_{1}}$ and the variance $\mu^{2} = \overline{(E_{1} - \overline{E_{1}})^{2}}$. Hence, there is an exact conservation of the adiabatic invariant whenever $\mu^{2} = 0$ because the system would have evolved from a single energy shell to another. In the remainder of this section, we will compare numerical results of $\mu^{2}$ and $W_{exc}$ for both systems mentioned previously.
 
%
%

\begin{figure}
\includegraphics[scale = 0.40]{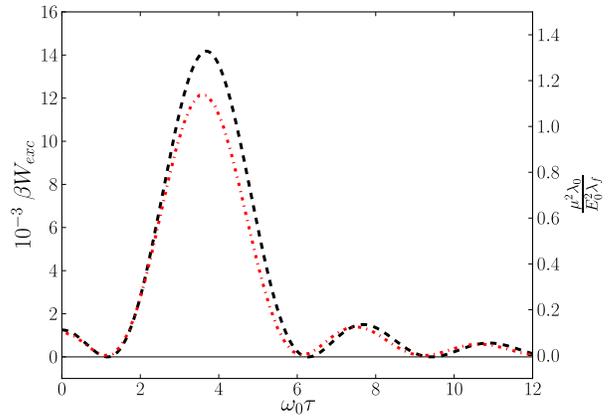} 
\caption{(Color online) Comparison between the analytical prediction of $W_{exc}$ (dashed line) given by (\ref{excess_work}) and the numerical calculation of $\mu^{2}$ for the harmonic oscillator (\ref{ohhamil}) using the protocol $g(t)$ = $(t-t_{0})/\tau + \sin(2\pi (t-t_{0})/\tau)$. We fixed $\delta\lambda/\lambda_{0} = 0.1$ and used $10^{6}$ initial conditions.}
\label{adiabatic_sin2x} 
\end{figure}

We see in Fig.~\ref{inv_adiabatic_harm} a clear agreement between the behaviors of $W_{exc}$ and $\mu^{2}$ for the harmonic oscillator. This result was obtained using a linear protocol. Furthermore, it is shown in Fig.~\ref{adiabatic_sin2x} that agreement is also very good for protocols of the family (\ref{gen_protocol}).  These results suggest that the finite-time zeros of $W_{exc}$ imply the conservation of $\Omega$. Conversely, we would like to have a proof that every time $\Omega$ is conserved in finite time, $W_{exc}$ vanishes. For the moment we only have numerical evidence that when the Hamiltonian (\ref{ohhamil}) is driven by the family of protocols (\ref{gen_protocol}), $\Omega$ is conserved whenever $W_{exc}$ vanishes in finite time. We have observed this no matter the initial energy shell we start.

\begin{figure}
\includegraphics[scale = 0.40]{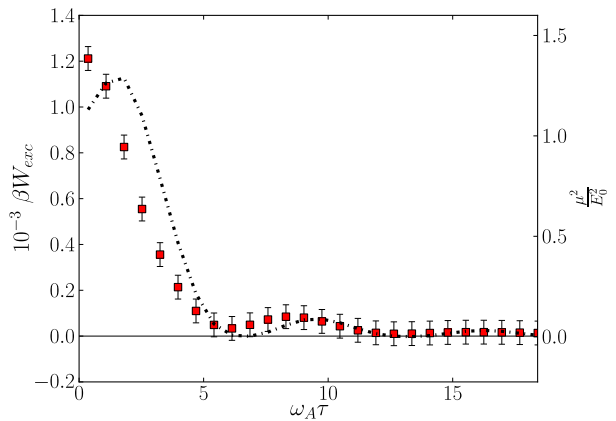} 
\caption{(Color online) Comparison between numerical calculations of $W_{exc}$ (squares) and $\mu^{2}$ (dashed line) for the anharmonic oscillator (\ref{quartic}) using the linear protocol. We fixed $\delta\lambda/\lambda_{0} = 0.1$ and used $10^{6}$ initial conditions.}
\label{adiabatic_quartic} 
\end{figure}

\begin{figure}
\includegraphics[scale = 0.40]{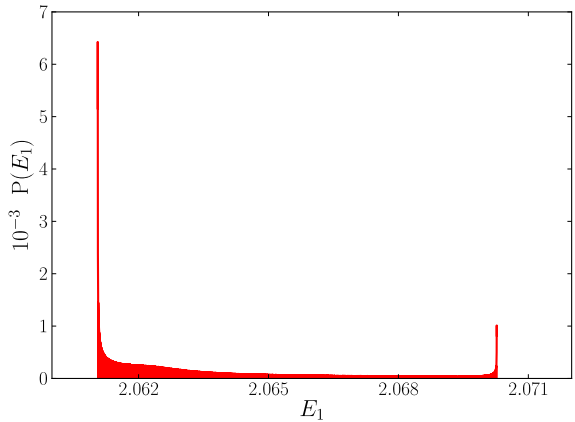} 
\caption{(Color online) Energy distribution at the end of the linear protocol for system (\ref{quartic}). We have chosen $\omega_{A}\tau$ such that $\mu^{2}$ has its first zero at this value (see Fig.~\ref{adiabatic_quartic}). Initial conditions were sampled from a single energy shell.}
\label{distribution_quartic} 
\end{figure}

The connection between the adiabatic invariant and thermodynamic work was not mentioned in previous works about optimal paths in thermally isolated systems. Figure \ref{adiabatic_quartic} suggests that such relation also exists for the anharmonic potential (\ref{quartic}). However, the zeros of $\mu^{2}$ do not imply the conservation of $\Omega$ in this case because Eqs.~(\ref{energy}) do not apply \cite{robnik,robnik3}, i.e., the vanishing of $\mu^{2}$ does not imply that all higher-order moments vanish too. This can be verified numerically, constructing the distribution of $E_{1}$ for the values of $\tau$ where $\mu^{2}$ vanishes. Figure \ref{distribution_quartic} shows that this distribution does not have a single peak, indicating that at the end of the protocol there is not just one single value of $\Omega$.

%

\section{Discussion \label{sec:discus}}

As mentioned before, the existence of finite-time processes leading to $W_{exc}=0$ was first reported in Ref.~\cite{seifert} for thermally isolated systems. There, the authors claim the existence of highly degenerate protocols for which the work performed is equal to the quasistatic value. In the regime described by our linear response approach, we were able not only to confirm the existence of such protocols but to show that they must obey a certain symmetry. According to Sec. \ref{sec:excoh}, the optimal finite-time protocols we found are composed of two parts: a linear protocol plus a function which is \textit{odd} with respect to time reversal. This symmetry embraces a much larger class of protocols than those of Ref.~\cite{seifert}. On the other hand, we have shown that these optimal protocols never lead to $W_{exc}=0$ for one-dimensional anharmonic oscillators. Instead, $W_{exc}$ has minima when such protocols are performed. The problem of $W_{exc}=0$ in finite time for nonlinear systems was addressed in Ref.~\cite{seifert} and has remained inconclusive. Unfortunately, our contribution to this problem is very restricted: we have shown that the family (\ref{gen_protocol}) does not lead to finite-time zeros of $W_{exc}$. Thus, our results do not exclude the existence of other families of optimal protocols.

The fact that $W_{exc}$, as a function of $\tau$, can have finite-time zeros or minima contradicts the usual intuition of a monotonic decay as the quasistatic limit is approached. At a first glance, this may be wrongly taken as an exclusive feature of systems with few degrees of freedom. However, Eq.~(\ref{excess_work}) tells us that what really matters is the behavior of the relaxation function. Therefore, we can infer from the analysis of the simple models presented here that this apparently peculiar thermodynamic behavior might also show up in the thermodynamic limit. This is the great advantage of our phenomenological approach. Indeed, it is well known that Eq.~(\ref{exponential_relaxation_func}) can describe the decay of correlations of a large class of macroscopic systems \citep{kubo2}. Although we leave for a future work the study of more complex systems, we would like to briefly outline how the physics of Eq.~(\ref{excess_work}) provides approximate solutions of the optimization problem in this case. If the relaxation function is known from computer simulations, then Eq.~(\ref{excess_work}) tells us that we must find a $\dot{g}(u)$ such that the surface generated by $\Psi_{0}(\tau(u-u')) \dot{g}(u) \dot{g}(u')$ has the minimal volume inside the integration domain. In other words, from the knowledge of $\Psi_{0}(t)$, it is possible to find approximate optimal solutions by geometric inspection.

Another aspect of the nonmonotonic decay of $W_{exc}$ is worth mentioning. As the switching time goes to zero, our results approach the value $\beta(W_{jp} - W_{qs})$ no matter the protocol we use. The quantity $W_{jp}$ represents the work performed when $\lambda$ is suddenly switched from $\lambda_{0}$ to $\lambda_{f}$ and its value is simply $\langle V(\lambda_{f}) - V(\lambda_{0})\rangle_{0}$, where $V(\lambda)$ is the potential energy. Thereby, this is the fastest protocol we can perform. One would expect then that $W_{exc}$ is maximum when $\tau\to 0$. However, this is not what we have observed. According to Figs.~\ref{sin_2x}, \ref{a_negative}, and \ref{fig:wexc_quartic}, there are indeed finite-time peaks of $W_{exc}$ whose values are much larger than $W_{exc}(\tau\to 0)$.

We also want to point out that definitions (\ref{eq_z}) and (\ref{expza}) are essentially the \textit{rapidity} parameter appearing in the study of quantum work distributions of the thermally isolated harmonic oscillator \cite{deffner,seifert,talkner}. From what was shown here, we believe that $z_{A}$ probably plays an important role in the statistics of quantum work of anharmonic oscillators \cite{obinna}. 

As a last remark, we want to mention the relation between the excess work and the conservation of the adiabatic invariant. Our results suggest that, in general, finite-time zeros (or minima) of $W_{exc}$ imply the conservation (or almost conservation) of $\Omega$. This relation deserves a more careful analysis due to its potential usefulness in the search for optimal paths. Besides, this has interesting implications to the adiabatic switching method proposed by Watanabe and Reinhardt \cite{watanabe} to estimate entropy and free-energy differences: whenever $\Omega$ is almost conserved in finite time, then this method will also provide good estimates in finite time. As discussed by the authors in Ref.~\cite{watanabe}, the problem of course is how to find, in general, the switching protocol that does the job. Our approach suggests that although the existence of such optimal paths depends very much on the dynamics of the system, it could be inferred from the behavior of the corresponding relaxation function (see Fig. \ref{exp_relaxation}).

\section{Conclusions \label{sec:conclu}}

In summary, we have shown that within linear response, there are highly degenerate protocols which lead to finite-time zeros or minima of the so-called excess work, $W_{exc}$, on a thermally isolated system. This quantity was defined as the amount of energy the external agent has to pump into the system in addition to the quasistatic work. Therefore, every time the excess work vanishes in finite time, the total work is equal to the quasistatic value. According to our approach, the family of optimal protocols must be composed of a linear part plus a function which is odd with respect to time reversal. 

Our analytical and numerical results have shown a counterintuitive behavior of the excess work as a function of the switching time, namely, a nonmonotonic decay as the process becomes slower. Although obtained for small systems, we claim that this behavior exists in macroscopic systems as well. Our argument relies on the expression for the excess work based on the relaxation function. We have shown that, for weak enough decay of correlations, this effect must be present no matter the size of the system. In other words, the only requirement is that the driving force $\partial H/\partial\lambda$ has a sufficiently oscillatory autocorrelation function.

Finally, the relation between finite-time zeros (or minima) of $W_{exc}$ and the conservation of the adiabatic invariant $\Omega$ suggests that there may exist an interesting and useful connection between optimal finite-time processes and shortcuts to adiabaticity \cite{jarz2,deffner3}.

\begin{acknowledgments}
Both authors thank C. Jarzynski for his hospitality during their visit to the University of Maryland, where most of this work was developed. It is also a pleasure to thank S. Deffner and M. de Koning for enriching discussions and Y. Suba\c{s}i and R. Freitas for useful comments and suggestions about the manuscript. T.A. acknowledges financial support from the Physics Institute of the Universidade Estadual de Campinas and CNPq (Brazil), Project No. 134296/2013-3. M.B. acknowledges financial support from FAPESP (Brazil), Project No. 2012/07429-0.
\end{acknowledgments}

\appendix
\section{Adiabatic invariant \label{sec:apC}}

In this Appendix, we obtain the adiabatic invariant $\Omega(E, \lambda)$ for the systems considered previously. Since it is the area enclosed by the energy shell \citep{Fasano}, $\Omega(E,\lambda)$ can be calculated as follows:
\begin{eqnarray}
\Omega(E,\lambda) &=& \int dx\,dp\,\Theta(E - \mathcal{H}(x,p;\lambda))\nonumber\\
& = & \int_{0}^{E}d\mathcal{H}\int_{0}^{2\pi}d\theta\, \mathcal{J}(\theta, \mathcal{H})\,,
\end{eqnarray}
where $\mathcal{J}$ is the Jacobian of the transformation $(x, p)\to(\theta, \mathcal{H})$. 

\subsection{Harmonic Oscillator}

For the Hamiltonian (\ref{ohhamil}), the transformation mentioned before reads 
\begin{equation}
p = \sqrt{2\mathcal{H}} \cos{(\theta)}\,, \ \ \ \ \ \ x = \sqrt{2\mathcal{H}/\lambda} \sin{(\theta)}\,,
\end{equation}
and its Jacobian is given by
\begin{equation}
\mathcal{J}(\theta,\mathcal{H}) = \begin{vmatrix}
\sqrt{2\mathcal{H}} \cos(\theta) &  \sin(\theta)/\sqrt{2\mathcal{H}} \\ 
-\sin(\theta) \sqrt{\frac{2\mathcal{H}}{\lambda}} & \cos(\theta)/\sqrt{2\mathcal{H}\lambda}
\end{vmatrix} = \dfrac{1}{\sqrt{\lambda}}\,.
\end{equation}
Hence, the adiabatic invariant reads
\begin{eqnarray}\label{adinvoh}
\Omega(E,\lambda) = \dfrac{2\pi E}{\sqrt{\lambda}}\,.
\end{eqnarray}

\subsection{Anharmonic Oscillator I}

Considering now the Hamiltonian (\ref{quartic}), the phase-space parametrization is given by
\begin{equation}
p = \sqrt{2\mathcal{H}} \sin{(\theta)}\,, \ \ \ \ \ \ x = \left(\dfrac{2\mathcal{H}}{\lambda}\right)^{1/4} \cos^{1/2}{(\theta)}\,,
\end{equation}
where $0 \leq \theta \leq \pi/2$ and its Jacobian reads
\begin{eqnarray}
\mathcal{J}(\theta,\mathcal{H}) &=& \begin{vmatrix}
\sqrt{2\mathcal{H}} \cos(\theta) & \sin(\theta)/\sqrt{2\mathcal{H}} \\ 
-\frac{1}{2} (\frac{2 \mathcal{H}}{\lambda})^{1/4} \frac{\sin(\theta)}{\cos^{1/2}(\theta)} &  \frac{\mathcal{H}^{-3/4} \cos^{1/2}(\theta)}{2^{7/4} \ \lambda^{1/4}}
\end{vmatrix} \nonumber \\
&=& \dfrac{\mathcal{H}^{-1/4}}{2^{5/4} \lambda^{1/4}} [\cos^{3/2}(\theta) + \sin^{2}(\theta) \cos^{-1/2}(\theta)]\,. \nonumber \\
\end{eqnarray}
Performing the integrals, we obtain
\begin{equation}\label{adinvoq}
\Omega(E,\lambda) = \dfrac{2^{13/4} \, K(1/2)}{3\, \lambda^{1/4}} E^{3/4}\,,
\end{equation}
where $K(m)$ is the incomplete elliptic integral of the first kind given by
\begin{equation}
K(m) = \int_{0}^{\pi/2}\frac{d\phi}{\sqrt{1-m \sin^{2}{(\phi)}}}\,.
\end{equation}

\subsection{Anharmonic Oscillator II}

The phase-space parametrization for Hamiltonian (\ref{sixth}) is given by
\begin{equation}
p = \sqrt{2\mathcal{H}} \sin{(\theta)}\,, \ \ \ \ \ \ x = \left(\dfrac{6\mathcal{H}}{\lambda}\right)^{1/6} \cos^{1/3}{(\theta)}\,,
\end{equation}
where $0 \leq \theta \leq \pi/2$. After calculating the Jacobian, 
\begin{eqnarray}
\mathcal{J}(\theta,\mathcal{H}) &=& \begin{vmatrix}
\sqrt{2\mathcal{H}} \cos(\theta) & \sin(\theta)/\sqrt{2\mathcal{H}} \\ 
-\frac{2^{1/6} \mathcal{H}^{1/6}}{3^{5/6} \lambda^{1/6}} \frac{\sin(\theta)}{\cos^{2/3}(\theta)} &  \frac{\mathcal{H}^{-5/6}}{6^{5/6} \lambda^{1/6}}\cos^{1/3}(\theta)
\end{vmatrix} \nonumber \\
&=&  \dfrac{\mathcal{H}^{-1/3}}{2^{1/3}3^{5/6}\lambda_{0}^{1/6}} \left[ \sin^{2}(\theta)\cos^{-2/3}(\theta) + \cos^{4/3}(\theta) \right]\,, \nonumber \\
\end{eqnarray}
we obtain
\begin{equation}\label{adinvs}
\Omega(E,\lambda) = \dfrac{2^{2/3} 3^{7/6} \sqrt{\pi}}{\lambda^{1/6}} \frac{\Gamma(7/6)}{\Gamma(2/3)} E^{2/3} \,,
\end{equation}
where $\Gamma(x)$ is the gamma function defined as
\begin{equation}
\Gamma(x) = \int_{0}^{\infty}dt\,t^{x-1} e^{-t}\,.
\end{equation}
%

\section{Calculation of quasistatic work \label{sec:apB}}

In this Appendix, we derive exact analytical expressions for the quasistatic work. This is the quantity we have to subtract from the numerical value of the mean work $W$ to obtain the excess work $W_{exc}$. Since the system is thermally isolated during its time evolution, we have from the first law of thermodynamics that $W = \Delta U$, where $U$ is the internal energy. Hence, in the quasistatic limit, 
\begin{equation}
W_{qs} = \Delta U = \langle E^{ad}_{f} \rangle - \langle E_{i} \rangle\,.
\end{equation}
where $E^{ad}_{f}$ is the energy obtained via the conservation of the adiabatic invariant as a function of the initial energy, $E_{i}$, and the initial and final values of the control parameter $\lambda$. The brackets $\langle . \rangle$ denote an average on a Boltzmann-Gibbs distribution since the system was initially in equilibrium with a heat bath. Besides, this distribution is taken with $\lambda = \lambda_{i}$, which is the initial value of $\lambda$.

\subsection{Harmonic Oscillator}

The adiabatic invariant of the harmonic oscillator (\ref{ohhamil}) is given by Eq.~(\ref{adinvoh}). Thus, for a quasistatic process, the conservation of $\Omega(E,\lambda)$ yields
\begin{equation}
E^{ad}_{f} = E_{i} \left( \frac{\lambda_{f}}{\lambda_{i}}\right)^{1/2}\,,
\end{equation}
which implies
\begin{eqnarray}\label{wqsoh}
W_{qs} = \dfrac{1}{\beta} \left[ \left(\dfrac{\lambda_{f}}{\lambda_{i}} \right)^{1/2} - 1\right] \,.
\end{eqnarray}

\subsection{Anharmonic Oscillator I}

Considering now a quasistatic process performed on the anharmonic oscillator (\ref{quartic}), the conservation of $\Omega(E,\lambda)$ given by (\ref{adinvoq}) yields
\begin{equation}
E^{ad}_{f} = E_{i} \left( \frac{\lambda_{f}}{\lambda_{i}}\right)^{1/3}\,,
\end{equation}
which implies
\begin{equation}
W_{qs} = \dfrac{3}{4\beta} \left[ \left(\dfrac{\lambda_{f}}{\lambda_{i}} \right)^{1/3} - 1\right]\,.
\end{equation}

\subsection{Anharmonic Oscillator II}

Finally, for the anharmonic oscillator (\ref{sixth}), the conservation of its adiabatic invariant (\ref{adinvs}) along a quasistatic process provides
\begin{equation}
E^{ad}_{f} = E_{i} \left( \frac{\lambda_{f}}{\lambda_{i}}\right)^{1/4}\,, 
\end{equation}
which implies
\begin{equation}
W_{qs} = \dfrac{2}{3\beta} \left[ \left(\dfrac{\lambda_{f}}{\lambda_{0}} \right)^{1/4} - 1\right]\,.
\end{equation}

\subsection{Linear Response expression of $W_{qs}$}

We can now compare the linear response expression for the quasistatic work,
\begin{equation}\label{apB:wexc}
W_{qs}^{LR} = \delta\lambda \left\langle\dfrac{\partial \mathcal{H}}{\partial \lambda} \right\rangle_{0} - \dfrac{(\delta \lambda)^{2}}{2} \tilde{\Psi}_{0}(0)\,,
\end{equation} 
with the exact results derived in Appendix \ref{sec:apC}.

For the harmonic oscillator (\ref{ohhamil}), we have
\begin{equation}
\left\langle \frac{\partial\mathcal{H}}{\partial\lambda} \right\rangle_{0} = \left\langle \frac{x^{2}}{2} \right\rangle_{0} = \frac{1}{2\beta\lambda_{0}}\,,
\end{equation}
and
\begin{eqnarray}
\tilde{\Psi}_{0} &=& \Psi_{0}(0) - \chi_{0}^{\infty} \nonumber\\
&=& \frac{\beta}{2}\left( \left\langle \frac{x^{4}}{4} \right\rangle_{0} - \left\langle \frac{x^{2}}{2} \right\rangle_{0}^{2} \right) 
= \frac{1}{4\beta\lambda_{0}^{2}}\,,
\end{eqnarray}
since $\chi_{0}^{\infty} = 0$.

Hence, Eq.~(\ref{apB:wexc}) reads
\begin{equation}\label{wqslr}
W_{qs}^{LR} = \dfrac{1}{\beta} \left[ \dfrac{\delta \lambda}{2\lambda_{0}} - \dfrac{1}{8} \left( \dfrac{\delta\lambda}{\lambda_{0}}\right)^{2}\right]\,.
\end{equation}

The expansion of the exact result (\ref{wqsoh}) for $\delta\lambda/\lambda_{0} \ll 1$ reads
\begin{eqnarray}
W_{qs} &=& \dfrac{1}{\beta} \left[ \left(\dfrac{ \lambda_{f}}{\lambda_{0}} \right)^{1/2} - 1\right] = \dfrac{1}{\beta} \left[ \left( 1 + \dfrac{\delta\lambda}{\lambda_{0}} \right)^{1/2} - 1\right] \nonumber \\
&= & \dfrac{1}{\beta} \left[ \dfrac{\delta \lambda}{2\lambda_{0}} - \dfrac{1}{8} \left( \dfrac{\delta\lambda}{\lambda_{0}}\right)^{2}\right] + O(3)\,,
\end{eqnarray}
which is equal to (\ref{wqslr}) up to second order in $\delta\lambda/\lambda_{0}$.

%
%

%

\end{document}